\documentclass[a4paper,fleqn,usenatbib]{mnras}

\usepackage{mathptmx}
\usepackage[T1]{fontenc}
\usepackage{ae,aecompl}

\usepackage{graphicx}	% Including figure files
\usepackage{amsmath}	% Advanced maths commands
\usepackage{amssymb}	% Extra maths symbols

\title[Observations of three cool ZZ~Ceti stars]{Wandering near the red edge: photometric observations of three cool ZZ~Ceti stars
}

\author[Zs. Bogn\'ar et al.]{Zs. Bogn\'ar,$^{1}$\thanks{E-mail: bognar.zsofia@csfk.mta.hu (Zs.B.)}
M. Papar\'o,$^{1}$
\'A. S\'odor,$^{1}$
D.~I. Jenei,$^{3}$
Cs. Kalup,$^{1,2}$
E. Bertone,$^{4}$
\newauthor
M. Chavez-Dagostino,$^{4}$
M.~H. Montgomery,$^{5}$
\'A. Gy\H orffy,$^{3}$
L. Moln\'ar,$^{1,6}$
H. Oll\'e,$^{3}$
\newauthor
P.~I. P\'apics,$^{7}$
E. Plachy,$^{1,6}$
E. Vereb\'elyi,$^{1}$
\\
$^{1}$Konkoly Observatory, MTA Research Centre for Astronomy and Earth Sciences, Konkoly Thege Mikl\'os \'ut 15-17, H--1121 Budapest\\
$^{2}$E\"otv\"os University, Department of Astronomy, Pf. 32, 1518, Budapest, Hungary\\
$^{3}$Visiting astronomer at Konkoly Observatory\\
$^{4}$Instituto Nacional de Astrof{\'\i}sica, \'Optica y Electr\'onica, Luis Enrique Erro 1, CP 72840, Tonantzintla, Puebla, Mexico\\
$^{5}$University of Texas at Austin, 2515 Speedway, Stop C1400, Austin, TX 78712-1205, USA\\
$^{6}$MTA CSFK Lend\"ulet Near-Field Cosmology Research Group, Konkoly Observatory, MTA Research Centre for Astronomy and \\Earth Sciences, Konkoly Thege Mikl\'os \'ut 15-17, H--1121 Budapest\\
$^{7}$Instituut voor Sterrenkunde, KU Leuven, Celestijnenlaan 200D, B-3001 Leuven, Belgium\\
}

\date{Accepted XXX. Received YYY; in original form ZZZ}

\pubyear{2018}

\begin{document}
\label{firstpage}
\pagerange{\pageref{firstpage}--\pageref{lastpage}}
\maketitle

\begin{abstract}
%This is a simple template for authors to write new MNRAS papers.
%The abstract should briefly describe the aims, methods, and main results of the paper.
%It should be a single paragraph not more than 250 words (200 words for Letters).
%No references should appear in the abstract.

We summarize our findings on three cool ZZ~Ceti type pulsating white dwarfs. We determined eight independent modes in HS~0733+4119, of which seven are new findings. For GD~154, we detected two new eigenmodes, and the recurrence of the pulsational behaviour first observed in 1977. We discuss that GD~154 does not only vary its pulsations between a multiperiodic and a quasi-monoperiodic phase, but there are also differences between the relative amplitudes of the near-subharmonics observed in the latter phase. In the complex pulsator, Ross~808, we compared the pre- and post Whole Earth Telescope campaign measurements, and determined two new frequencies besides the ones observed during the campaign. Studying these stars can contribute to better understanding of pulsations close to the empirical ZZ~Ceti red edge. All three targets are in that regime of the ZZ~Ceti instability strip where short-term amplitude variations or even outbursts are likely to occur, which are not well-understood theoretically.

\end{abstract}


%%%%%%%%%%%%%%%%%%%%%%%%%%%%%%%%%%%%%%%%%%%%%%%%%%

%%%%%%%%%%%%%%%%% BODY OF PAPER %%%%%%%%%%%%%%%%%%

\section{Introduction}

%All papers should start with an Introduction section, which sets the work
%in context, cites relevant earlier studies in the field by \citet{Others2013},
%and describes the problem the authors aim to solve \citep[e.g.][]{Author2012}.

White dwarfs represent the final evolutionary stage of initially low- to intermediate-mass stars ($\sim97$\% of all stars in the Universe). By the investigation of their interiors, we can find the clue to the previous steps of stellar evolution, use them as cosmic laboratories, and measure the age of the stellar population they are belonging to. 

Some of them show light variations on time scales of minutes. We can find them in well-defined regions of the Hertzsprung--Russell-diagram, and they can be classified in three major groups: DAV (ZZ~Ceti), DBV (V777 Her) and GW Vir stars. The most populous one is that of ZZ~Ceti, $\sim80$\% of the known pulsating white dwarfs belong to this group. Contrary to DBVs and GW Vir stars, ZZ~Ceti stars has a thin hydrogen layer above the helium one. These stars have some hydrogen left in spite of the effective mass loss processes that occured on the asymptotic giant branch.

ZZ\,Ceti stars are short-period and low-amplitude pulsators with 10\,500--13\,000\,K effective temperatures. Their light variations are caused by non-radial \textit{g}-mode pulsations with periods in the 100--1500\,s range, typically with $\sim$mmag amplitudes. It has been known for decades, that ZZ~Ceti stars close to the blue edge of the instability strip show different pulsational behaviour than their cooler siblings in the middle of the ZZ~Ceti instability domain or close to its red edge. While we can detect a few small-amplitude eigenmodes at the hottest ZZ~Cetis, the cooler variables are more likely to show a larger number of pulsation modes with higher amplitudes. There are also differences in their temporal pulsational behaviour, as the cool ZZ~Cetis often show frequencies with variable amplitudes and phases in short time scales (days-weeks), while the pulsation of hotter ones are more stable \citep{1999ApJ...511..904G}. We also note the observations of the so-called `outburst' events, which means recurring increases in the stellar flux (up to 15\%) in cool ZZ~Ceti stars (see e.g. \citealt{2017ASPC..509..303B}).

The pulsation periods of variable stars are sensitive to the global stellar structure, rotation, and internal chemical stratification. However, for the in-depth asteroseismic investigations, the precise determination of as many eigenmodes as possible is needed. This motivated us in the selection of our targets: we focused on cool ZZ~Ceti stars with the possibility of showing rich frequency spectra and amplitude variations or even outbursts. 
Assuming that the global structure of the star does not change on time-scales as short as the amplitudes of the excited frequencies, we observe a subset of the possible eigenmodes during one observing run. However, we can complement the list of significant frequencies with observations at different epochs. That is, more observations provide a more complete sampling of the pulsation modes for asteroseismic analysis.

The outburst phenomenon might be in connection with the cessation of pulsations at the empirical red edge of the ZZ~Ceti instability strip \citep{2015ApJ...810L...5H}, which mechanism is still an open question from the theoretical point of view. It is also not well-understood, how and why exactly the pulsation amplitudes vary in short time-scales if these phenomena are not results of insufficient frequency resolution of the datasets. Possible explanations are interactions of pulsation and convection (e.g. \citealt{2010ApJ...716...84M}), and resonant mode couplings (e.g. in \citealt{2016A&A...585A..22Z}).

According to the ground-based observations and the measurements with the \textit{Kepler} space telescope, we can distinguish five stages of the cooling of the ZZ~Ceti stars
\citep{2017ApJS..232...23H}. (1) The stars show short-period (100--300\,s), low-amplitude ($\sim$1\,mmag) low-radial-order pulsations at the blue edge of the instability strip. (2) At a few hundred degrees cooler stage, the periods are still short, but the pulsation amplitudes rise up to $\sim$5\,mmag, and some modes are stable enough to investigate evolutionary period changes. (3) In the middle of the instability strip, we find the pulsators with the highest amplitudes, often showing several nonlinear combination frequencies in their Fourier spectrum. In some cases, short-term amplitude and frequency variations emerge. (4) As the stars cool further, they may show irregularly recurring outbursts. (5) We can find the longest pulsation periods amongst the coolest ZZ~Ceti stars, and as we reach the edge of the instability domain, their amplitudes decrease. For a short explanation why the cooler ZZ~Cetis show longer pulsation periods, see e.g. \citet{1993BaltA...2..407C}.

We selected three stars for observations, situated at the cool part of the instability strip:
HS~0733+4119 (Sect.~\ref{sect:hs}), an object with known pulsation modes only by its discovery runs; GD~154 (Sect.~\ref{sect:gd}), a relatively well-studied star that exhibited different pulsation behaviours in the past; and Ross~808 (Sect.~\ref{sect:ross}), a star that had been targeted by an international campaign earlier, but we collected a significant amount of additional data before and after the campaign's observing window, giving us the chance to compare its pulsation behaviour at different epochs.

In general, the more precise pulsation frequencies lead to more reasonable astreroseismic modeling results. Our goal is to investigate as many stars as possible for this purpose. This way we can get a more detailed picture on the ZZ~Ceti stars as a whole group. Moreover, besides the space observations, ground-based measurements obtained at different epochs also play an important role in the investigations of temporal pulsation variations over different time-scales. All three of our targets are in that regime of the ZZ~Ceti instability strip where short-term amplitude variations or even outbursts are likely to occur (see Sect.~\ref{sect:summary}), which are not well-understood theoretically. Studying these stars can contribute to better understanding of pulsations close to the empirical ZZ~Ceti red edge.

%All of our targets are located in the regime of the ZZ~Ceti instability strip where outbursts are expected, see Sect.~\ref{sect:summary}. In this paper, we summarize our findings on these three objects.

\section{Observations, data reduction, and analysis}
\label{sect:redu}

We performed the observations with the 1-m Ritchey--Chr\'etien--Coud\'e telescope located at the Piszk\'estet\H o mountain station of Konkoly Observatory, Hungary. The measurements were made either with a Princeton Instruments VersArray:1300B back-illuminated CCD camera or with an Andor iXon+888 EMCDD in white light. We used either 10 or 30\,s exposure times, depending on the weather conditions. The read-out time was $\sim 3$\,s for the VersArray:1300B camera, while it was negligible thanks to the frame transfer mode in the case of the iXon+888 EMCDD.
In the case of GD~154, we utilized additional data from other observatories, described in Sect.~\ref{sect:gd}.

Raw data frames were treated the standard way utilizing \textsc{iraf}\footnote{\textsc{iraf} is distributed by the National Optical Astronomy Observatories, which are operated by the Association of Universities for Research in Astronomy, Inc., under cooperative agreement with the National Science Foundation.} tasks: we performed bias, dark and flat corrections before the aperture photometry of field stars. After photometry, we fitted low-order (second- or third-order) polynomials to the resulting light curves, correcting for long-period atmospheric and instrumental effects. This procedure did not affect the known frequency domain of pulsating ZZ~Ceti stars, however, this detrending could make the detection of any possible outburst events difficult or even impossible. The comparison stars for the differential photometry were checked for variability and instrumental effects. Then, we converted the observational times of every data point to barycentric Julian dates in barycentric dynamical time ($\mathrm{BJD_{TDB}}$) using the applet of \citet{2010PASP..122..935E}\footnote{http://astroutils.astronomy.ohio-state.edu/time/utc2bjd.html}.    

We analysed the daily measurements with the command-line light curve fitting program \textsc{LCfit} \citep{2012KOTN...15....1S}. \textsc{LCfit} has linear (amplitudes and phases) and nonlinear (amplitudes, phases and frequencies) least-squares fitting capabilities, utilizing an implementation of the Levenberg-Marquardt least-squares fitting algorithm. The program can handle unequally spaced and gapped datasets and is easily scriptable.

We performed standard Fourier analyses on the weekly-monthly and whole datasets with the photometry modules of the Frequency Analysis and Mode Identification for Asteroseismology (\textsc{famias}) software package \citep{2008CoAst.155...17Z}. We accepted a frequency peak as significant if its amplitude reached the 4 signal-to-noise ratio (S/N), where noise level was determined by the average Fourier amplitude in a 1700\,$\mu$Hz radius vicinity of the peak in question. We chose this smoothing radius somewhat arbitrarily, but taking into account the structure of the central peak of the window function, the spectral resolution, and the rate of the background spectral-noise variation.

\section{HS\,0733+4119}
\label{sect:hs}

HS\,0733+4119 ($g=15.8$\,mag, $\alpha_{2000}=07^{\mathrm h}37^{\mathrm m}08^{\mathrm s}$, $\delta_{2000}=+41^{\mathrm d}12^{\mathrm m}28^{\mathrm s}$) is a known pulsator. Its light variations were detected by \citet{2007ASPC..372..583V}. They determined three pulsation periods at 747.4, 656.2, and 468.8\,s. No further time-series observations of the star can be found in the literature.

We observed HS\,0733+4119 on 12 nights between January and April, 2017. Table~\ref{tabl:loghs} shows the journal of observations.

\begin{table}
\centering
\caption{Log of observations on HS\,0733+4119. `Exp' is the integration time used, \textit{N} is the number of data points, and $\delta T$ is the length of the data sets including gaps. Weekly observations are denoted by `a,b,c,d' letters in parentheses.}
\label{tabl:loghs}
\begin{tabular}{lrccrr}
\hline
Run & UT Date & Start time & Exp. & \textit{N} & $\delta T$ \\
 & (2017) & (BJD-2\,450\,000) & (s) &  & (h) \\
\hline
01(a) & Jan 19 & 7773.305 & 10 & 3026 & 8.58\\
02(a) & Jan 22 & 7776.411 & 10 & 2352 & 6.68\\
03(b) & Jan 26 & 7780.235 & 10 & 3358 & 10.60\\
04(b) & Jan 27 & 7781.213 & 10 & 6181 & 11.22\\
05(b) & Jan 28 & 7782.214 & 10 & 7203 & 11.02\\
06(c) & Feb 19 & 7804.415 & 30 & 274 & 2.31\\
07(c) & Feb 21 & 7806.358 & 10 & 1759 & 6.04\\
08(c) & Feb 23 & 7808.338 & 10 & 486 & 1.51\\
09(c) & Feb 25 & 7810.226 & 10 & 2703 & 8.21\\
10(c) & Feb 26 & 7811.473 & 10 & 1250 & 3.79\\
11(d) & Apr 01 & 7845.277 & 10 & 1282 & 3.69\\
12(d) & Apr 02 & 7846.264 & 10 & 858 & 2.77\\
\multicolumn{2}{l}{Total:} & & \multicolumn{2}{r}{30\,732} & 76.42\\
\hline
\end{tabular}
\end{table}

\begin{figure}
\centering
\includegraphics[width=0.47\textwidth]{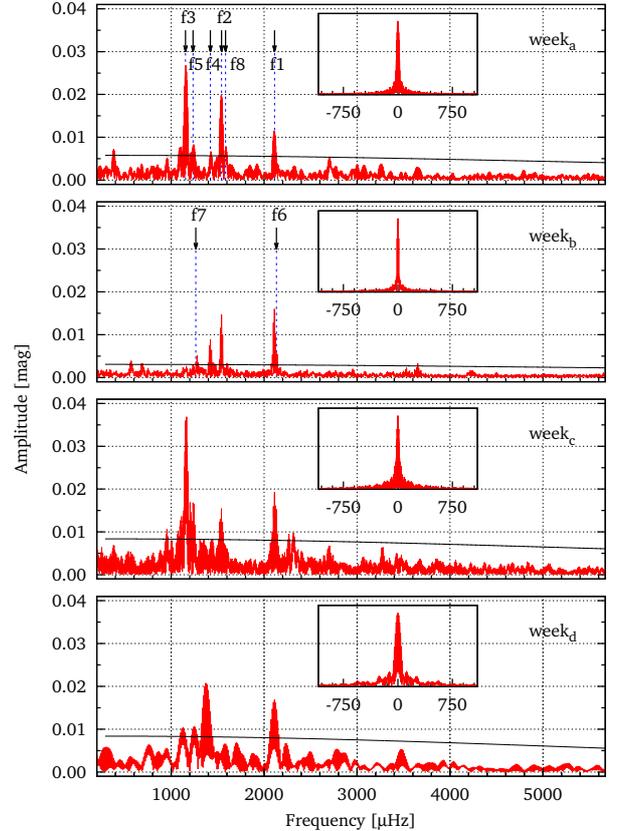}
\caption{Fourier transforms of the weekly datasets obtained on HS\,0733+4119. Note the remarkable amplitude variations from week to week. The longest observational runs were performed during the second week. The window functions are given in the insets. We also denote the 4$\langle {\rm A}\rangle$ significance levels (black lines), and mark the frequencies $f_1-f_8$ of Table~\ref{tabl:hsfrek} (cf. Fig.~\ref{fig:hsftwhole}), respectively.}{\label{fig:weeklyhs}}
\end{figure}

We analysed independently both the daily datasets and the weekly data. Finally, we performed the Fourier analysis of the whole dataset. To be rigorous in our frequency determination, we accepted a frequency if it was found to be significant at least on one night, in one weekly dataset and in the whole data as well. We summarized the finally accepted 14 frequencies in Table~\ref{tabl:hsfrek}. Note that there are additional significant peaks in the Fourier transform of the whole dataset, mostly closely spaced ones to the frequencies listed in Table~\ref{tabl:hsfrek}, but these most likely emerge as results of short-term amplitude or phase variations. Figure~\ref{fig:weeklyhs} clearly presents the remarkable amplitude variations from one week to another. 

We also checked the closely spaced peaks to any kind of regularities (e.g. doublets, triplets), which may emerge as results of rotational splitting of frequencies, but we found no convincing recurring frequency spacings. 

As Table~\ref{tabl:hsfrek} shows, we identified six combination peaks out of the 14 frequencies. The remaining eight peaks seem to be independent pulsation modes. Comparing our period list with the one presented by \citet{2007ASPC..372..583V}, $f_6$ is a common period (469.1 vs. 468.8\,s), but the status of the other two periods of \citet{2007ASPC..372..583V} at 656.2 and 747.4\,s is questionable. These either emerge in our dataset at 648.8 and 703.3\,s ($f_2$ and $f_4$), respectively, or considering the relatively large separations between our and the \citet{2007ASPC..372..583V} measurements, these two periods published in 2007 are also independent ones, and were not excited to an observable level at the time of our measurements.  

\begin{figure*}
\centering
\includegraphics[width=0.95\textwidth]{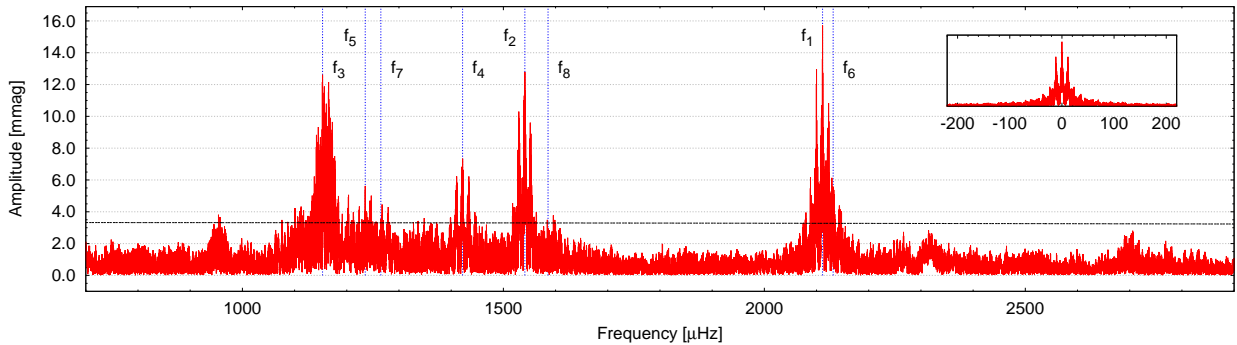}
\caption{Fourier transform of the whole dataset obtained on HS\,0733+4119. The frequencies which can be regarded as independent pulsation frequencies are marked with blue dashed lines (cf. Table~\ref{tabl:hsfrek}). The window function is given in the inset, while black line denotes the 4$\langle {\rm A}\rangle$ significance level.}{\label{fig:hsftwhole}}
\end{figure*}

Appendices~\ref{app:hslc} and \ref{app:hsft} shows the daily light curves and their Fourier transforms, respectively. We plotted the Fourier transform of the whole dataset in Fig~\ref{fig:hsftwhole} and denoted the eight independent frequencies.    

\begin{table}
\centering
\caption{HS\,0733+4119: list of finally accepted frequencies by the whole dataset. The frequencies are listed in the order of the pre-whitening procedure. The errors are formal uncertainties, and are estimated to be $\sim 0.3$\,mmag in amplitudes. We discuss the influence of the 1\,d$^{-1}$ aliasing on the frequency determination in Sect.~\ref{sect:summary}.}
\label{tabl:hsfrek}
%\tiny
\begin{tabular}{lrrrrr}
\hline
 & \multicolumn{1}{c}{\textit{f}} & \multicolumn{1}{c}{\textit{P}} & Ampl. & Phase & Comment\\
 & \multicolumn{1}{c}{[$\mu$Hz]} & \multicolumn{1}{c}{[s]} & [mmag] & [$2\pi$] & \\
\hline
$f_1$ &	2111.580(2) & 473.6 & 15.6 & 0.66 \\
$f_2$ & 1541.240(2) & 648.8 & 13.1 & 0.45 \\
$f_3$ & 1153.553(2) & 866.9 & 12.4 & 0.24 \\
$f_4$ & 1421.949(3) & 703.3 & 7.5 & 0.32 \\
$f_5$ & 1235.346(5) & 809.5 & 4.9 & 0.92 \\
$f_6$ &	2131.800(6) & 469.1 & 4.7 & 0.65 \\
%$f_7$ &	1126.786(14) & 887.5 & 1.8 & 0.76 \\
$f_7$ & 1265.321(8) & 790.3 & 3.3 & 0.58 \\
$f_8$ & 1585.624(7) & 630.7 & 3.9 & 0.01 \\
$f_9$ & 570.690(7) & 1752.3 & 3.4 & 0.45 & $f_1-f_2$\\
$f_{10}$ & 954.179(7) & 1048.0 & 3.8 & 0.79 & $\sim f_1-f_3$\\
$f_{11}$ & 379.830(9) & 2632.8 & 2.8 & 0.67 & $\sim f_2-f_3$\\
$f_{12}$ & 2705.945(21) & 369.6 & 2.7 & 0.24 & $\sim f_2+f_3$\\
$f_{13}$ & 2316.714(10) & 431.6 & 2.7 & 0.57 & $\sim 2f_3$\\
$f_{14}$ & 3652.825(10) & 273.8 & 2.6 & 0.22 & $f_1+f_2$\\
\hline
\end{tabular}
\end{table}

\section{GD\,154}
\label{sect:gd}

GD\,154 ($V=15.3$\,mag, $\alpha_{2000}=13^{\mathrm h}09^{\mathrm m}58^{\mathrm s}$, $\delta_{2000}=+35^{\mathrm d}09^{\mathrm m}47^{\mathrm s}$) is a well-studied ZZ~Ceti star; it was the target of both multisite campaigns and single-site observations. It was found to be a variable in 1977, with one dominant frequency at 843\,$\mu$Hz and its harmonic and near-subharmonic peaks \citep{1978ApJ...220..614R}. Many years later, in 1991, it was selected to be a target of a Whole Earth Telescope (WET; \citealt{1990ApJ...361..309N}) campaign. At that time, it showed three independent modes at 842.8, 918.6, and 2484.1\,$\mu$Hz with their rotationally split components and combination frequencies, but peaks were not found around subharmonic frequencies \citep{1996A&A...314..182P}. During a two-site campaign organized in 2004, new excited modes were detected at 786.5, 885.4, and 1677.7\,$\mu$Hz \citep{2005ASPC..334..577H}. At last, results of extended single-site observations from 2006 were published by \citet{2013MNRAS.432..598P}. They determined six normal modes, two of which, at 807.62 and 861.56\,$\mu$Hz were new detections. 

We continued the observations of this star and obtained data on eight nights in 2012. Table~\ref{tabl:loggd} shows the journal of these observations, while Appendices~\ref{app:gdlc} and \ref{app:gdsp} present the daily light curves and their Fourier spectra, respectively.

\begin{figure*}
\centering
\includegraphics[width=0.95\textwidth]{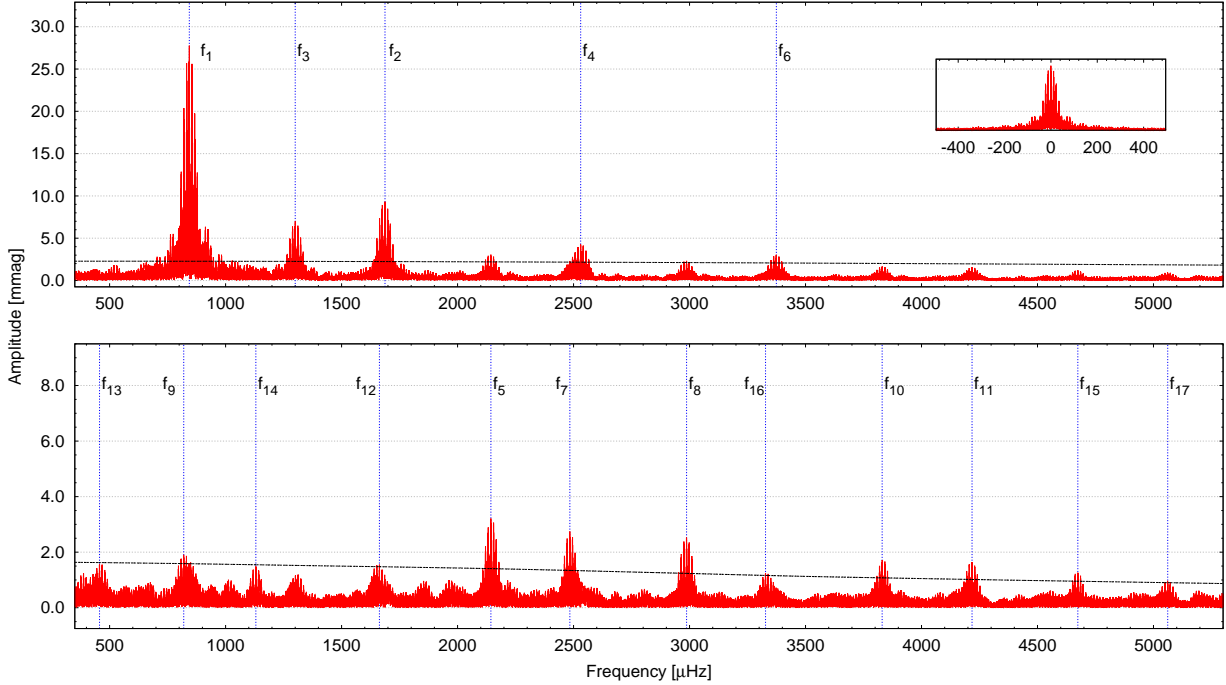}
\caption{Fourier transform of the whole 2012 dataset obtained on GD\,154. The frequencies of Table~\ref{tabl:gdfrek} are denoted by blue dashed lines. For the better visibility of the frequencies, we plot the Fourier spectrum after pre-whitening with $f_1$, $f_2$, $f_3$, $f_4$ and $f_6$ in the bottom panel. Black lines denote the 4$\langle {\rm A}\rangle$ significance levels. The window function is given in the inset.}{\label{fig:gdftwhole}}
\end{figure*}

As the frequency analyses of both the daily and the whole datasets show, we observed a pulsation behaviour very similar to what was seen in 1977. We list the frequencies obtained by the whole dataset in Table~\ref{tabl:gdfrek}. One frequency ($f_1 = 843.6\,\mu$Hz) dominates the Fourier transforms, while a series of harmonic ($n \times f_1$, where $n=2,3,4,5,6$) and near-subharmonic peaks ($m \times f_1$, where $m=0.54, 1.54, 2.54, 3.54, 4.54, 5.54$) are clearly visible. However, the light variations are not monoperiodic, that is, at least two more independent modes can be determined ($f_7=2483.9\,\mu$Hz and $f_{14}=1130.5\,\mu$Hz). The status of two peaks, $f_9=819.6\,\mu$Hz and $f_{12}=1663.1\,\mu$Hz is questionable. Both $f_9$ and $f_{12}$ are similarly close ($\delta f \sim 2$d$^{-1}$) to another peak, $f_1$ and $f_2$, respectively, while $f_2$ (at $1687.1\,\mu$Hz) is the first harmonic of $f_1$. We see two possibilities here: (1) $f_9$ is an independent mode and $f_{12}=f_1+f_9$, (2) the presence of $f_9$ is due to some kind of amplitude or phase modulation of $f_1$ and it is reflected at $f_2$ as the peak of $f_{12}$. We plot the Fourier transform of the whole dataset in Fig~\ref{fig:gdftwhole}.

%We summarized the normal modes from different epochs for comparison in Table~\ref{tabl:gdcomp}, and plot the Fourier transform of the whole dataset in Fig~\ref{fig:gdftwhole}.

We also had the chance to follow the changes in the pulsation behaviour of this star examining measurements from three additional epochs (2005, 2007, and 2011). We obtained data on GD\,154 at the McDonald Observatory (McDO) with the 2.1-meter Otto Struve Telescope through a BG40 filter in 2005 and 2007, respectively. The star was observed on three nights in 2005, and on seven nights in 2007, respectively. We also collected additional observations from 2011 with the 2.1-m telescope at the Guillermo Haro Astrophysical Observatory (OAGH), Mexico, utilizing a $B$ filter. The star was observed on three consecutive nights. We included the journal of these observations in Table~\ref{tabl:loggd}, while the figures of Appendix~\ref{app:gdlcusa} show the corresponding light curves. We used frame transfer mode in the case of the McDO observations, while the read-out times were 7\,s for the OAGH measurements.

The raw data frames of the OAGH observations were reduced the same way as the Konkoly measurements, see Sect.\ref{sect:redu}. The observations performed at McDonald Observatory were treated similarly, the difference is that in this case we utilised the flux values instead of the magnitudes. Thus, the resulting normalized light curves show light variations represented as fractional intensity in modulation intensity (mi) units, see the top panels of Appendix~\ref{app:gdlcusa}. To convert the corresponding milli-modulation amplitudes (mma) of the Fourier transforms of the McDO light curves to milli-magnitude (mmag) values used in all the other cases, we applied the relationship as follows: \mbox{1 mma $= 1/1.086$ mmag.} 

We summarized the eigenmodes of the datasets from 1977 to 2012 in Table~\ref{tabl:gdcomp}. We detected two new normal modes at 798.3 (2007) and 1130.5\,$\mu$Hz (2012), respectively.

\subsection{Near-subharmonics in the datasets}

The star's pulsation behaviour was similar to what was observed in 1977 and what we measured in 2007, 2011, and 2012, that is, one dominant mode and its harmonic and near-subharmonic peaks dominated the light variations. However, we note a remarkable difference: in 2011, the second highest amplitude peak was at $\sim1.5f_1$ and not at the first harmonic of the dominant peak.
Furthermore, there are no detectable peaks at the near-subharmonic frequency domains in the 2005 dataset, leastways at $\sim$5\,mmag detection limit. That is, the star does not only vary its pulsational behaviour between a `simple' multiperiodic (see \citealt{1996A&A...314..182P}, \citealt{2005ASPC..334..577H} and \citealt{2013MNRAS.432..598P}) and a quasi monoperiodic phase with harmonic and near-subharmonic peaks, but there are differences between the visibility of the near-subharmonics in this latter phase, too (see the relative amplitudes listed in Table~\ref{tabl:gdampl}). This reminds us to the finding of \citet{1978ApJ...220..614R}, when the authors reported that the $\sim1.5f_1$ peak became the dominant on the last night of their observations. This pulsational-behaviour change of GD\,154 raises the question: what physical mechanism can cause such rapid nonlinear response in the star?

Similarly to the 2012 observations, both in the 2007 and 2011 datasets the near-subharmonic peaks appeared at $m \times f_1$, where $m=1.54, 2.54, 3.54$ (2007) or $m=1.54, 2.54$ (2011). \citet{1978ApJ...220..614R} reported $1.52$, $2.53$, and $3.53\,F$ frequencies. This means that there is a clear and consistent shift from the real subharmonic frequencies expected at $n/2 \times f_x$, where $n$ is an odd integer.

The near-subharmonic frequencies resemble the pulsational behaviour observed in other white dwarf variables, e.g., in the DBV star PG\,1351+489 \citep{1988A&A...196L..13G} and in the ZZ~Cetis G191-16 \citep{1989A&A...215L..17V} and BPM\,31594 \citep{1992MNRAS.258..415O}. 
In BPM\,31594, a similar ($1.54 \pm n$)$f$ near-subharmonic frequency set was detected. However, the authors emphasized that considering the observations of these near-subharmonics as signs of period doubling is wrong: there is a clear frequency shift between the expected real \mbox{($1.5 \pm n$)$f$} subharmonic and the observed ($1.54 \pm n$)$f$ near-subharmonic peaks.
Nevertheless, this phenomenon could still indicate some non-linear interaction or coupling between certain modes in the star. 

\begin{figure}
\centering
\includegraphics[width=0.43\textwidth]{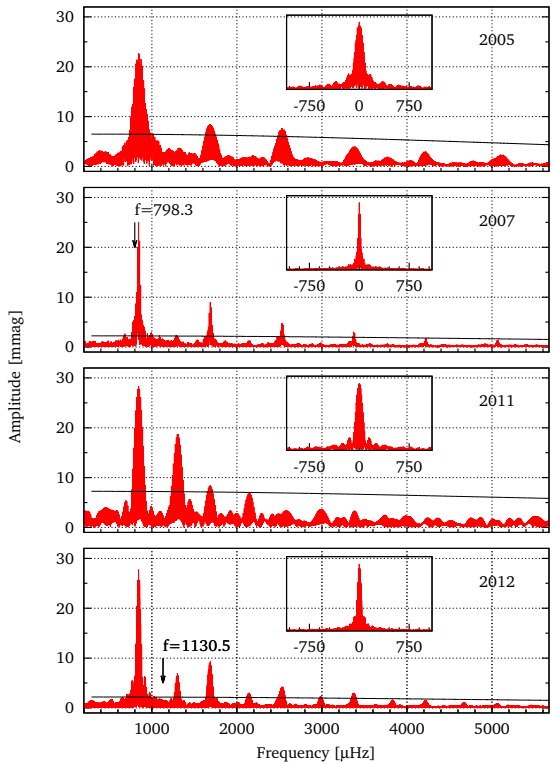}
\caption{Fourier transforms of the whole datasets of the 2005, 2007, 2011 and 2012 measurements of GD\,154. Note the remarkable amplitude variations of the near-subharmonic peaks from one epoch to another (cf. Table~\ref{tabl:gdampl}). Black lines denote the 4$\langle {\rm A}\rangle$ significance levels, while the window functions are given in the insets. We also mark the two newly detected eigenmodes (cf. Table~\ref{tabl:gdcomp}). Note that these peaks are clearly visible and significant after pre-whitening with the larger-amplitude peaks.}{\label{fig:compgd}}
\end{figure}

Figure~\ref{fig:compgd} shows the Fourier transforms of the 2005, 2007, and 2011 observations together with the 2012 measurement at Konkoly Observatory.

\begin{table}
\centering
\caption{Log of observations on GD\,154. `Exp' is the integration time used, \textit{N} is the number of data points and $\delta T$ is the length of the data sets including gaps.}
\label{tabl:loggd}
\begin{tabular}{lrccrr}
\hline
Run & UT Date & Start time & Exp. & \textit{N} & $\delta T$ \\
 &  & (BJD-2\,450\,000) & (s) &  & (h) \\
\hline
\multicolumn{6}{l}{Konkoly Obs. (2012)}\\
01 & Jan 27 & 5953.509 & 30 & 517 & 4.67\\
02 & Jan 28 & 5954.619 & 30 & 249 & 2.24\\
03 & Jan 30 & 5956.509 & 30 & 542 & 4.91\\
04 & Jan 31 & 5957.624 & 30 & 237 & 2.15\\
05 & Feb 01 & 5958.505 & 30 & 549 & 4.92\\
06 & Feb 20 & 5978.445 & 30 & 344 & 3.16\\
07 & Feb 21 & 5979.450 & 30 & 593 & 5.82\\
08 & Feb 22 & 5980.449 & 30 & 598 & 5.34\\
\multicolumn{2}{l}{Total:} & & \multicolumn{2}{r}{3\,629} & 33.21\\
\\
\multicolumn{6}{l}{McDonald Obs. (2005)}\\
01 & May 10 & 3500.620 & 10 & 733 & 2.03\\
02 & May 12 & 3502.611 & 10 & 393 & 1.10\\
03 & May 13 & 3503.756 & 10 & 948 & 2.64\\
\multicolumn{2}{l}{Total:} & & \multicolumn{2}{r}{2\,074} & 5.77\\
\\
\multicolumn{6}{l}{McDonald Obs. (2007)}\\
01 & Apr 20 & 4210.671 & 15 & 1083 & 4.61\\
02 & Apr 21 & 4211.597 & 15 & 2243 & 9.40\\
03 & Apr 22 & 4212.729 & 15 & 2380 & 6.22\\
04 & Apr 24 & 4214.597 & 15 & 2165 & 9.36\\
05 & Apr 25 & 4215.610 & 10 & 915 & 3.04\\
06 & Apr 26 & 4216.595 & 10 & 1003 & 2.81\\
07 & Apr 28 & 4218.611 & 10 & 906 & 2.51\\
\multicolumn{2}{l}{Total:} & & \multicolumn{2}{r}{10\,695} & 37.96\\
\\
\multicolumn{6}{l}{Guillermo Haro Astrophysical Obs. (2011)}\\
01 & Mar 09 & 5629.925 & 70 & 115 & 2.41\\
02 & Mar 10 & 5630.921 & 70 & 119 & 2.67\\
03 & Mar 11 & 5631.914 & 70 & 128 & 2.98\\
\multicolumn{2}{l}{Total:} & & \multicolumn{2}{r}{362} & 8.06\\
\hline
\end{tabular}
\end{table}

\begin{table}
\centering
\caption{GD\,154: frequencies determined by the whole dataset obtained in 2012. The frequencies are listed in the order of the pre-whitening procedure. The errors are formal uncertainties, and are estimated to be $\sim 0.2$\,mmag in amplitudes. We discuss the influence of the 1\,d$^{-1}$ aliasing on the frequency determination in Sect.~\ref{sect:summary}.}
\label{tabl:gdfrek}
%\tiny
\begin{tabular}{lrrrrr}
\hline
 & \multicolumn{1}{c}{\textit{f}} & \multicolumn{1}{c}{\textit{P}} & Ampl. & Phase & Comment\\
 & \multicolumn{1}{c}{[$\mu$Hz]} & \multicolumn{1}{c}{[s]} & [mmag] & [$2\pi$] & \\
\hline
$f_1$ &	843.565(1)	&	1185.4	&	27.5	&	0.10 &	indep. mode\\
$f_2$	&	1687.127(4)	&	592.7	&	8.9	&	0.02 &	2$f_1$\\
$f_3$	&	1300.112(4)	&	769.2	&	7.0	&	0.55 &	1.54$f_1$\\
$f_4$	&	2530.697(6)	&	395.1	&	4.5	&	0.58 &	3$f_1$\\
$f_5$	&	2143.691(9)	&	466.5	&	3.2	&	0.93 &	2.54$f_1$\\
$f_6$	&	3374.273(9)	&	296.4	&	3.0	&	0.16 &	4$f_1$\\
$f_7$	&	2483.949(10)	&	402.6	&	2.7	&	0.52 &	indep. mode\\
$f_8$	&	2987.270(11)	&	334.8	&	2.5	&	0.37 &	3.54$f_1$\\
$f_9$	&	819.564(15)	&	1220.2	&	2.1	&	0.74 &	indep. mode?\\
$f_{10}$	&	3830.358(16)	&	261.1	&	1.7	&	0.13 &	4.54$f_1$\\
$f_{11}$	&	4217.868(16)	&	237.1	&	1.7	&	0.02 &	5$f_1$\\
$f_{12}$	&	1663.104(16)	&	601.3	&	2.0	&	0.19 &	$f_1+f_9$?\\
$f_{13}$	&	456.481(19)	&	2190.7	&	1.5	&	0.22 &	0.54$f_1$\\
$f_{14}$	&	1130.461(19)	&	884.6	&	1.4	&	0.24 &	indep. mode\\
$f_{15}$	&	4673.908(22)	&	214.0	&	1.3	&	0.66 &	5.54$f_1$\\
$f_{16}$	&	3327.549(23)	&	300.5	&	1.2	&	0.63 &	$f_1+f_7$\\
$f_{17}$	&	5061.389(30)	&	197.6	&	0.9	&	0.90 &	6$f_1$\\
\hline
\end{tabular}
\end{table}

\begin{table}
\centering
\caption{GD\,154: amplitudes and relative amplitudes of the dominant frequency, its first harmonic and the 1.54$f_1$ peak, fitting the light curves with these frequencies only. Large variation in the amplitude of the near-subharmonic frequency can be detected.}
\label{tabl:gdampl}
%\tiny
\begin{tabular}{p{4mm}cccccc}
\hline
 & $A_{f_1}$ & $A_{1.54f_1}$ & $A_{2f_1}$ &  $A_{1.54f_1}/A_{f_1}$ & $A_{1.54f_1}/A_{2f_1}$ & $A_{2f_1}/A_{f_1}$ \\
 & \multicolumn{3}{c}{[mmag]} & \multicolumn{3}{c}{[\%]} \\
\hline
2005 & 23.1 & -- & 8.6 & -- & -- & 37.2\\
2007 & 25.0 & 2.6 & 9.1 & 10.4 & 28.6 & 36.4\\
2011 & 28.1 & 19.2 & 8.2 & 68.3 & 234.1 & 29.2\\
2012 & 27.7 & 7.1 & 9.3 & 25.6 & 76.3 & 33.6\\
\hline
\end{tabular}
\end{table}

%!!! ráírnám a panelekre az adatok forrását (távcső)

\begin{table*}
\centering
\caption{GD\,154: normal modes from different epochs. `H', `SH', `C' and `R' refer to that harmonic, near-subharmonic, combination (including harmonic) or rotationally split peaks were reported at the given dataset.}
\label{tabl:gdcomp}
\begin{tabular}{rrrrrrrrrrrrr}
\hline
Year & \multicolumn{10}{c}{Frequency} & Comm. & Reference\\
 & \multicolumn{10}{c}{($\mu$Hz)} & & \\
\hline
{\it 1977} & & & & {\it 843.1} & & & & & & & {\it H,SH} & \citet{1978ApJ...220..614R}\\
{\it 1991} & & & & {\it 842.8} & & & {\it 918.6} & & & {\it 2484.1} & {\it C,R} & Pfeiffer et al. (1996)\\
{\it 2004} & {\it 786.5} & & & & & {\it 885.4} & & & \it 1677.7 & \it 2484.3 & \it C,R & \citet{2005ASPC..334..577H}\\
2005 & & & & 843.2 & & 888.6 & & & & 2480.1 & H & this work: McDO \\
\it 2006 & & & \it 807.6 & \it 839.1 & \it 861.6 & \it 883.6 & \it 918.7 & & & \it 2484.1 & \it C,R & \citet{2013MNRAS.432..598P}\\
2007 & & 798.3 & & 843.3 & & & & & & 2483.8 & H,SH & this work: McDO\\
2011 & & & & 844.0 & & & & & & & H,SH & this work: OAGH, Mexico \\
2012 & & & & 843.6 & & & & 1130.5 & & 2483.9 & C,SH & this work: Konkoly Obs. \\
\hline
\end{tabular}
\end{table*}

\section{Ross\,808}
\label{sect:ross}

The light variations of Ross\,808 (hereafter R\,808; $V=14.4$\,mag, $\alpha_{2000}=16^{\mathrm h}01^{\mathrm m}23^{\mathrm s}$, $\delta_{2000}=+36^{\mathrm d}48^{\mathrm m}34^{\mathrm s}$) was discovered in 1975. At that time, one stable frequency was determined at around 1200\,$\mu$Hz, while the authors mentioned a complex, variable power spectrum of this star \citep{1976ApJ...205L.155M}. Decades later, \citet{2009MNRAS.396.1709C} presented new observations of R\,808. They identified eight frequencies at 926.7, 1045.5, 1101.8, 1139.1, 1255.8, 1342.1, 1955.8, and 2472.2\,$\mu$Hz. In the meantime, R\,808 was selected to be a target of a WET campaign performed in 2008 \citep{2009JPhCS.172a2067T}. This resulted in the detection of 28 frequencies, including linear combinations and closely spaced peaks, which may emerge as the results of rotational splitting of frequencies \citep{2009AIPC.1170..621B}.

\begin{figure}
\centering
\includegraphics[width=0.43\textwidth]{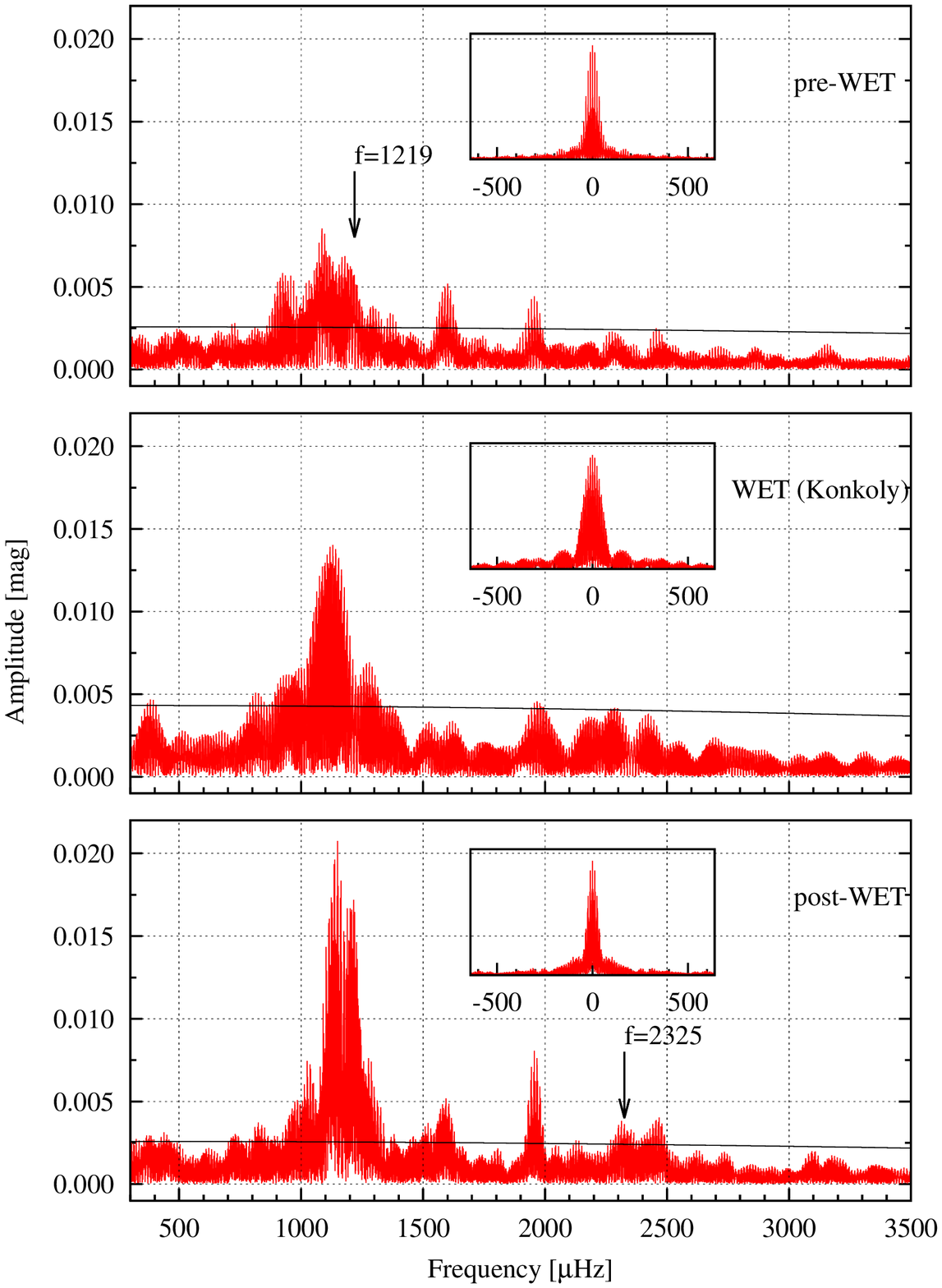}
\caption{R\,808: Fourier transforms of the pre- and post-WET measurements, and the observations obtained during the WET campaign. Black lines denote the 4$\langle {\rm A}\rangle$ significance levels, while the window functions are given in the insets. We also mark the two newly detected frequencies at 1219 and 2325\,$\mu$Hz.}{\label{fig:four808}}
\end{figure}

We observed R\,808 on 24 nights also in 2008. We joined the WET campaign mentioned above and observed in the frame of this multisite collaboration on four nights in April. Otherwise, most of our data were collected before (in January and in February) and after the WET observations (in May), see the details in Table~\ref{tabl:logr808}. The Fourier transforms of the pre- and post-WET observations, and the ones obtained during the WET campaign are plotted in Fig.~\ref{fig:four808}. Changes in the observed amplitudes and the complexity of the frequency structure in the $\sim$900--1400\,$\mu$Hz domain are clearly visible.

\begin{table} 
\centering
\caption{Log of observations on R\,808. `Exp' is the integration time used, \textit{N} is the number of data points and $\delta T$ is the length of the data sets including gaps.}
\label{tabl:logr808}
\begin{tabular}{lrccrr}
\hline
Run & UT Date & Start time & Exp. & \textit{N} & $\delta T$ \\
 & (2008) & (BJD-2\,450\,000) & (s) &  & (h) \\
\hline
01 & Jan 22 & 4487.566 & 30 & 150 & 1.49 \\
02 & Jan 23 & 4488.526 & 30 & 521 & 4.69 \\
03 & Feb 08 & 4504.542 & 10 & 898 & 3.23 \\
04 & Feb 09 & 4505.521 & 10 & 1192 & 4.09 \\
05 & Feb 10 & 4506.585 & 10 & 747 & 2.70 \\
06 & Feb 11 & 4507.595 & 10 & 695 & 2.50 \\
07 & Feb 12 & 4508.527 & 10 & 1170 & 4.18 \\
08 & Feb 13 & 4509.533 & 10 & 1123 & 3.98 \\
09 & Feb 15 & 4512.467 & 30 & 601 & 5.52 \\
10 & Feb 16 & 4513.466 & 30 & 601 & 5.49 \\
11 & Feb 18 & 4515.449 & 30 & 183 & 1.67 \\
12 & Feb 20 & 4516.511 & 30 & 490 & 4.46 \\
13 & Feb 20 & 4517.457 & 30 & 625 & 5.67 \\
14 & Apr 04 & 4560.512 & 10 & 731 & 2.80 \\
15 & Apr 14 & 4570.536 & 10 & 598 & 2.18 \\
16 & Apr 14 & 4571.473 & 10 & 837 & 3.30 \\
17 & Apr 17 & 4573.506 & 10 & 711 & 2.64 \\
18 & May 02 & 4589.432 & 10 & 328 & 1.18 \\
19 & May 03 & 4590.356 & 10 & 305 & 1.16 \\
20 & May 06 & 4593.419 & 10 & 1096 & 3.93 \\
21 & May 07 & 4594.420 & 10 & 1089 & 3.93 \\
22 & May 15 & 4602.378 & 10 & 1331 & 4.88 \\
23 & May 16 & 4603.342 & 10 & 1562 & 5.88 \\
24 & May 21 & 4608.301 & 10 & 1522 & 6.75 \\
\multicolumn{2}{l}{Total:} & & \multicolumn{2}{r}{19\,106} & 88.30\\
\hline
\end{tabular}
\end{table}

As the frequency content of R\,808 has become known for the duration of the WET campaign, it made us wonder how this may vary compared to the pre- and post-WET observations; did any new frequencies emerge with which we can complete the WET list of frequencies? For this purpose, we linearly fitted the pre- and post-WET datasets with the frequencies determined by the whole WET data presented by \citet{2009JPhCS.172a2067T}, and checked the frequency content of the residual Fourier spectra, searching for any new eigenmodes for R\,808. We also performed the independent Fourier analyses of the pre- and post-WET observations.
Our strategy was to accept a significant frequency in the Fourier transform as a newly detected one if it was at least $1$\,d$^{-1}$ far from the WET frequencies. Applying this criteria, we tried to avoid accepting a closely spaced peak to a WET mode, which may emerge as the result of non-stationary amplitudes and phases, or just the $1$\,d$^{-1}$ alias ambiguity of our single-site observations. Then we compared the results we obtained with these two methods, and searched for the common frequency solutions.      

This way we identified two more frequencies using the pre- and post-WET datasets. These represent new findings additionally to the WET frequencies presented by \citet{2009JPhCS.172a2067T}. These are at 1218.64(2)\,$\mu$Hz -- 821\,s (pre-WET observations) and at 2325.22(2)\,$\mu$Hz -- 430\,s (post-WET observations). Note that this latter peak is close to the first harmonic of F18 of \citet{2009JPhCS.172a2067T} (1162.48\,$\mu$Hz), however, we did not find this parent frequency analysing of our post-WET observations. Thus, there is a chance that this is only a coincidence caused by the large number of excited frequencies. The frequency values and their formal uncertainties are calculated from the averages of the frequencies derived by the two methods.

\section{Summary and Conclusions}
\label{sect:summary}

In this paper, we presented our findings on three cool ZZ~Ceti stars observed mainly at Konkoly Observatory. We focused on the frequency analyses of the targets and put these into context of the previous publications on the targets.

In the rarely observed HS\,0733+4119, we detected eight independent modes. With one exception, these modes seem to be new findings.

In the relatively well-studied GD\,154, we detected two new eigenmodes at 798.3 and 1130.5\,$\mu$Hz (cf. Table~\ref{tabl:gdcomp}). We also observed the recurrence of the pulsational behaviour first reported in 1977. Similarly to was found with this previous observation, one mode and its harmonic and near-subharmonic peaks dominated the light variations of the star in 2007, 2011 and 2012. We discussed the differences between its pulsational behaviour observed in different epochs; we found that the star does not simply alternate its pulsation behaviour between a multiperiodic and a quasi-monoperiodic fashion with harmonic and near-subharmonic peaks, but there are also differences between the relative amplitudes of the near-subharmonics in this latter phase at different times. The phenomenon that harmonic and
near-subharmonic peaks dominate the Fourier spectrum from time-to-time, while there are `simple' multiperiodic intervals when they disappear, shows the importance of nonlinear effects in the star's pulsation.

In Ross\,808, which is a pulsator rich in frequencies and which shows complex light variations, we compared the pre- and post-Whole-Earth-Telescope campaign measurements, and determined two new frequencies besides the ones observed during the campaign.

Considering the reliability of the frequency determinations of the presented ground-based observations, we concluded the followings. The largest uncertainties come from the presence of the 1\,d$^{-1}$ alias peaks in the Fourier spectra, that is, the real uncertainties can be $\pm$1\,d$^{-1}$ for the frequencies. However, in the case of HS\,0733+4119, we tested our frequency determination by the analyses of different data subsets (see Sect.~\ref{sect:hs}), identified combination terms, and found one common frequency with the previous observations. In GD\,154, we analysed datasets from different epochs and found the parent modes and their harmonic and near-subharmonic peaks at very similar frequencies. In the case of R\,808, we accepted the new frequencies by the analyses of data subsets with two different methods. These efforts and findings significantly lower the possibility that we confused a real pulsation frequency with a $\pm$1\,d$^{-1}$ alias.     

\begin{figure}
\centering
\includegraphics[width=0.49\textwidth]{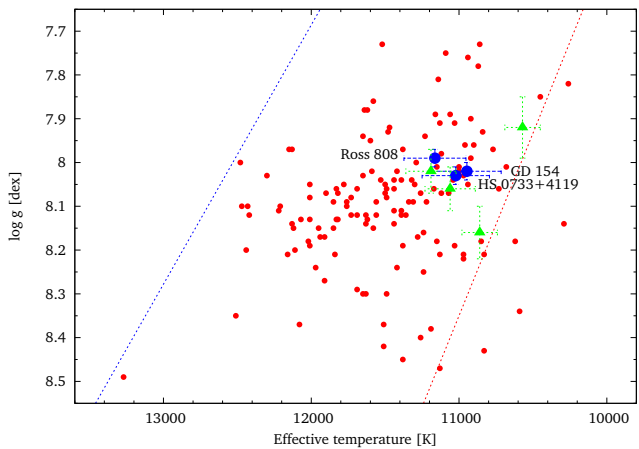}
\caption{The classical ZZ~Ceti instability strip with plots of the known ZZ~Ceti stars (red filled dots) including the stars presented in this paper (blue dots with errorbars) and the four outbursting ZZ~Cetis with published atmospheric parameters by \citet{2017ASPC..509..303B} (green triangles). Blue and red dashed lines denote the hot and cool boundaries of the instability strip, according to \citet{2015ApJ...809..148T}.}
{\label{fig:instab}}
\end{figure}

Figure~\ref{fig:instab} shows the classical ZZ~Ceti instability strip with all the known ZZ~Ceti stars plotted (red filled dots) including the stars presented in this paper (blue dots with errorbars). The atmospheric parameters of the ZZ~Cetis are from the database of \citet{2016IBVS.6184....1B} and from the Montreal White Dwarf Database (MWDD; \citealt{2017ASPC..509....3D})\footnote{http://dev.montrealwhitedwarfdatabase.org/home.html}. The pulsation properties of the three stars discussed in this paper are in good agreement with their location on the ZZ~Ceti instability strip: they show complex temporal light variations and also relatively large amplitudes, which all correspond to stars evolving from the middle of the instability domain towards the red edge of the instability strip.

However, considering the results of \textit{Kepler} observations of cool ZZ~Ceti stars, we can raise the question: do we see any signs of outbursts in the light curves? As \citet{2016ApJ...829...82B} and \citet{2017ASPC..509..303B} have already indicated, it is possible that we smooth the average brightness increases of the outburst events out with the polynomial fitting method used to correct for the long-period extinction and instrumental effects. However, we have the possibility to search for signs of changes in the stars' pulsational behaviour, for example, the amplitudes of the excited modes change during an outburst.

The best candidate for an outbursting star seems to be GD~154, which shows remarkable features in its light curve. First, as we have mentioned already, it has two pulsational states: a quasi-monoperiodic one with one dominant frequency and its harmonic and near-subharmonic peaks, and a `simple' multiperiodic state. This is definitely a dramatic change in the star's pulsational behaviour. However, note that considering the time scales, the star remains in the quasi-monoperiodic or the multiperiodic state much longer than the average duration of the outbursts (hours to about one day). 

Another possibility that we consider the light curve features presented by \citet{2013MNRAS.432..598P}, sect.~3.2.3. One conspicuous feature that we cannot fit adequately with the frequency solutions derived for the dataset at some high-amplitude, strongly non-sinusoidal intervals. These are about 0.1\,d long parts of the light curve. The conclusion in \citet{2013MNRAS.432..598P} was that it is possible that `some additional physical process is superimposed on the pulsation creating high-amplitude phases'.

Nevertheless, GD~154 and the other two stars are all in the domain of the instability strip where outbursts were detected by the \textit{Kepler} space telescope -- see Fig.~\ref{fig:instab}, where we plotted our three stars together with the four outbursting \textit{Kepler} objects with published effective temperatures and surface gravities. Fortunately, all three objects are on the list of short cadence targets proposed for \textit{TESS} (Transiting Exoplanet Survey Satellite; \citealt{2015JATIS...1a4003R}) observations. With those observations, we will have the opportunity to observe possible outburst phenomena by the uninterrupted measurements of the all-sky space project.

%\begin{figure}
%\centering
%\includegraphics[width=0.49\textwidth]{hrd_spec.eps}
%\caption{The stars presented in this paper (blue dots with errorbars) together with the four outbursting ZZ~Cetis with published atmospheric parameters by \citet{2017ASPC..509..303B} (green dots). Red dashed line denotes the cool boundary of the instability strip, according to \citet{2015ApJ...809..148T}.}
%{\label{fig:hrd2}}
%\end{figure}

\section*{Acknowledgements}
The authors thank the anonymous referee for the constructive comments and recommendations on the manuscript.
\'AS was supported by the J\'anos Bolyai Research Scholarship of the Hungarian Academy of Sciences, and he also acknowledges the financial support of the Hungarian NKFIH Grant K-113117. \'AS and ZsB acknowledge the financial support of the Hungarian NKFIH Grants K-115709 and K-119517. ZsB acknowledges the support provided from the National Research, Development and Innovation Fund of Hungary, financed under the PD\_17 funding scheme, project no. PD-123910. CsK acknowledges the financial support of the GINOP-2.3.2-15-2016-00003 and GINOP-2.3.2-15-2016-00033 grants of the Hungarian National Research, Development and Innovation Office. MC and EB would like to thanks financial support from CONACyT through
grants CB-2015-256961 and CB-2011-169554. LM and EP acknowledge the financial support of the Hungarian National Research, Development, and Innovation Office through the NKFIH Grants K-115709, PD-116175, and PD-121203; the LP2014-17, and LP2018-7/2018 Programs of the Hungarian Academy of Sciences, and they were also supported by the J\'anos Bolyai Research Scholarship of the Hungarian Academy of Sciences. This project has been supported by the Lend\"ulet grant LP2012-31 of the Hungarian Academy of Sciences.

%This work benefited a lot from the discussions within the SoFAR international team \footnote{\url{http://www.issi.unibe.ch/teams/sofar/}}
%supported by the International Space Science Institute (ISSI). 

%%%%%%%%%%%%%%%%%%%%%%%%%%%%%%%%%%%%%%%%%%%%%%%%%%



\bibliographystyle{mnras}
\bibliography{3stars} % if your bibtex file is called example.bib

%\begin{thebibliography}{99}
%\end{thebibliography}

%%%%%%%%%%%%%%%%%%%%%%%%%%%%%%%%%%%%%%%%%%%%%%%%%%

%%%%%%%%%%%%%%%%% APPENDICES %%%%%%%%%%%%%%%%%%%%%

\appendix

\section{}
%\label{app:novft}

\begin{figure*}
\centering
\includegraphics[width=0.7\textwidth, angle=270]{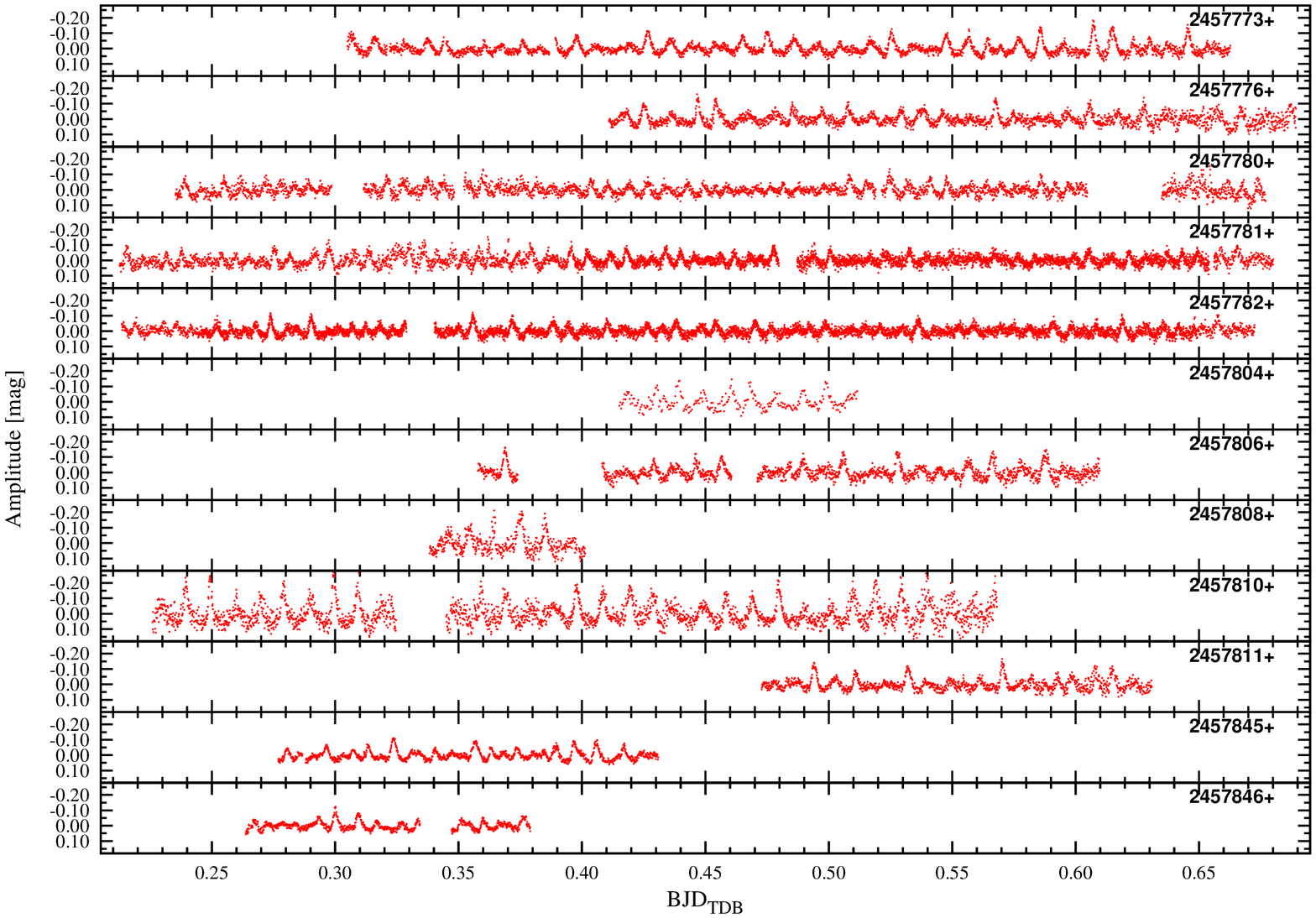}
\caption{Normalized differential light curves of the observations of HS\,0733+4119.}{\label{app:hslc}}
\end{figure*}

Figure~\ref{app:hslc}: Normalized differential light curves of the observations of HS\,0733+4119.

\section{}
%\label{app:novft}

\begin{figure*}
\centering
\includegraphics[width=0.75\textwidth]{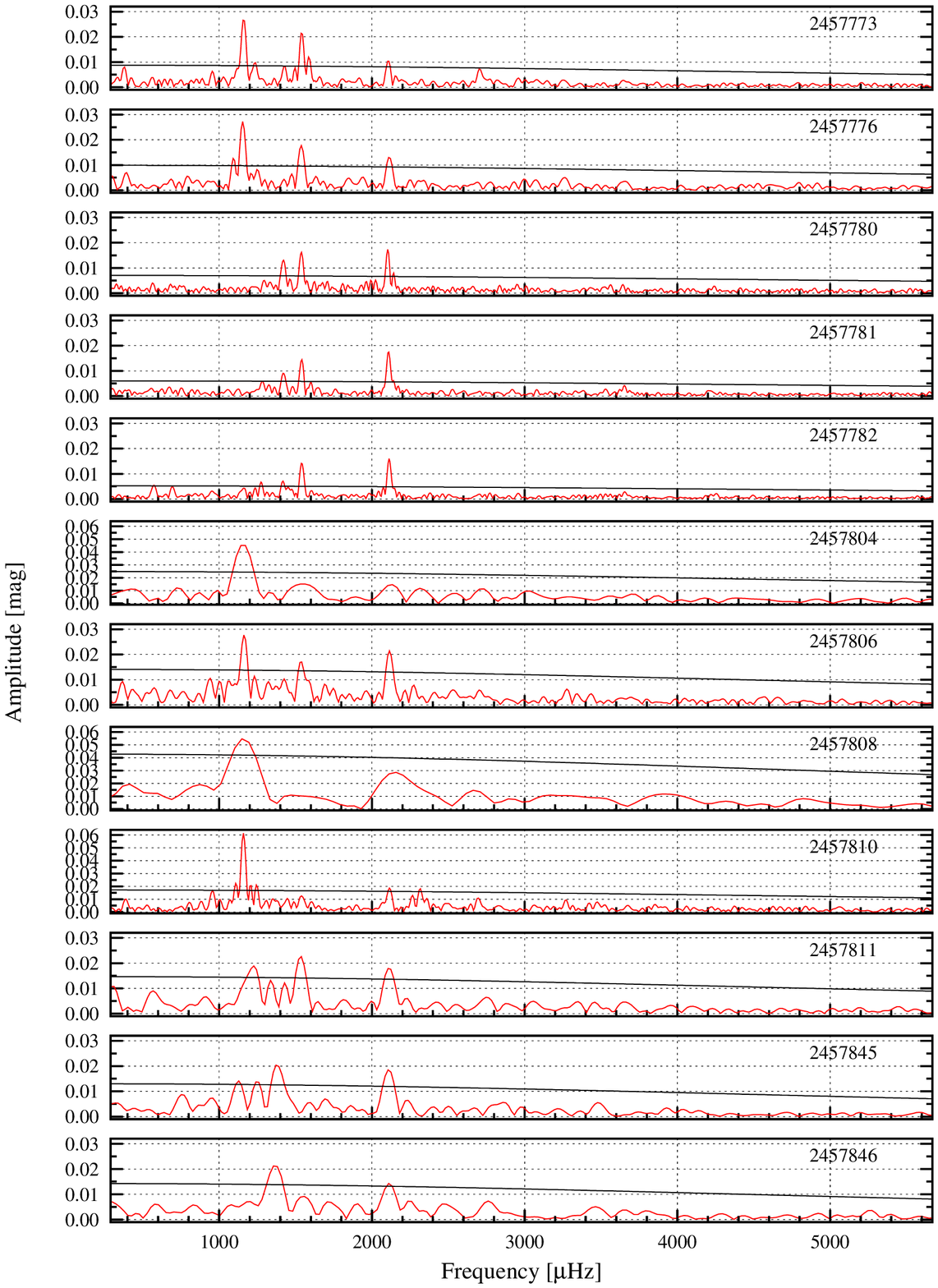}
\caption{Fourier transforms of the nightly observations of HS\,0733+4119. Black lines denote the 4$\langle {\rm A}\rangle$ significance levels.}{\label{app:hsft}}
\end{figure*}

Figure~\ref{app:hsft}: Fourier transforms of the nightly observations of HS\,0733+4119.

\section{}
%\label{app:novft}

\begin{figure*}
\centering
\includegraphics[width=0.7\textwidth, angle=270]{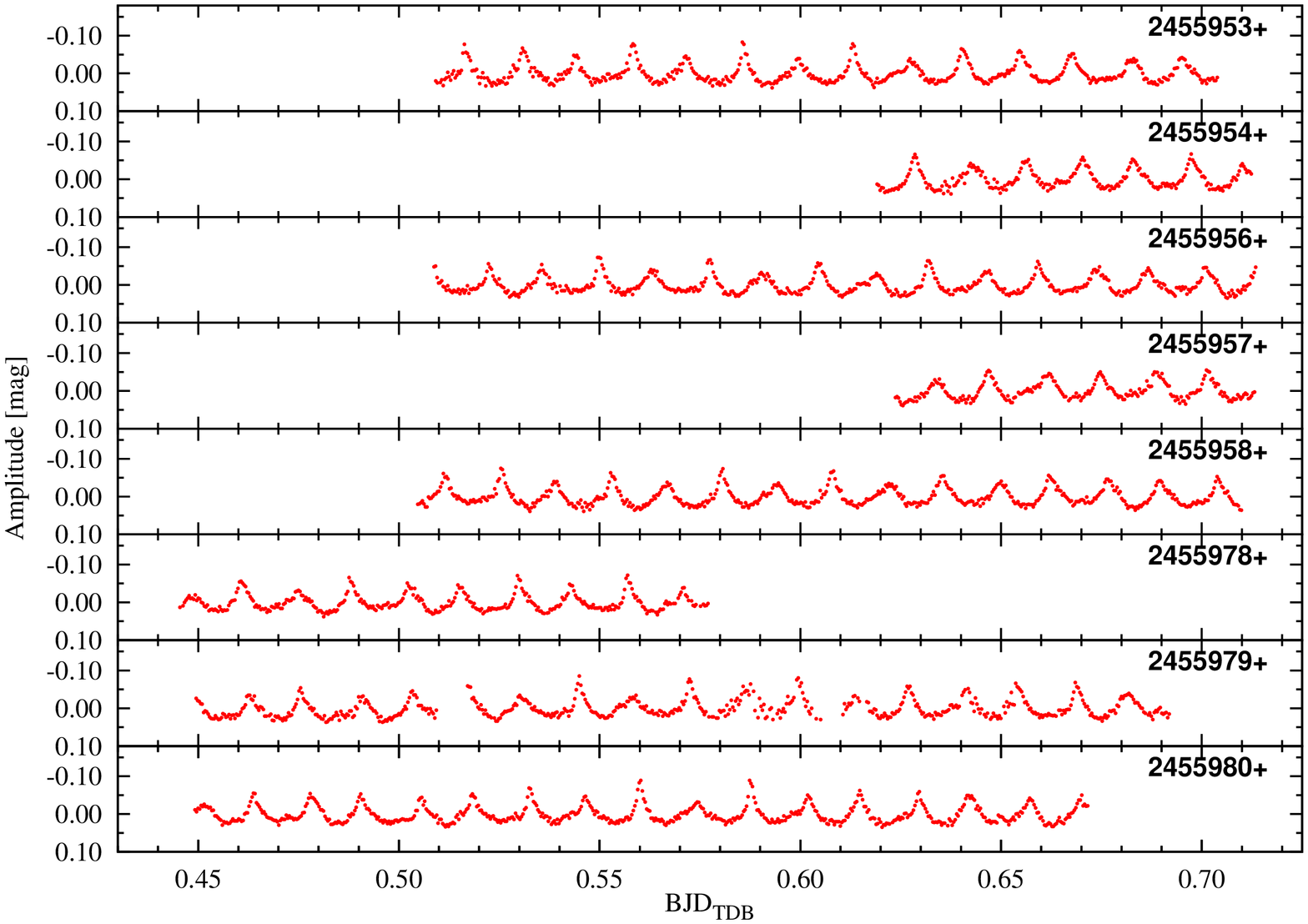}
\caption{Normalized differential light curves of the observations of GD\,154 at Konkoly Observatory.}{\label{app:gdlc}}
\end{figure*}

Figure~\ref{app:gdlc}: Normalized differential light curves of the observations of GD\,154 at Konkoly Observatory.

\section{}
%\label{app:wdft}

\begin{figure*}
\centering
\includegraphics[width=0.75\textwidth]{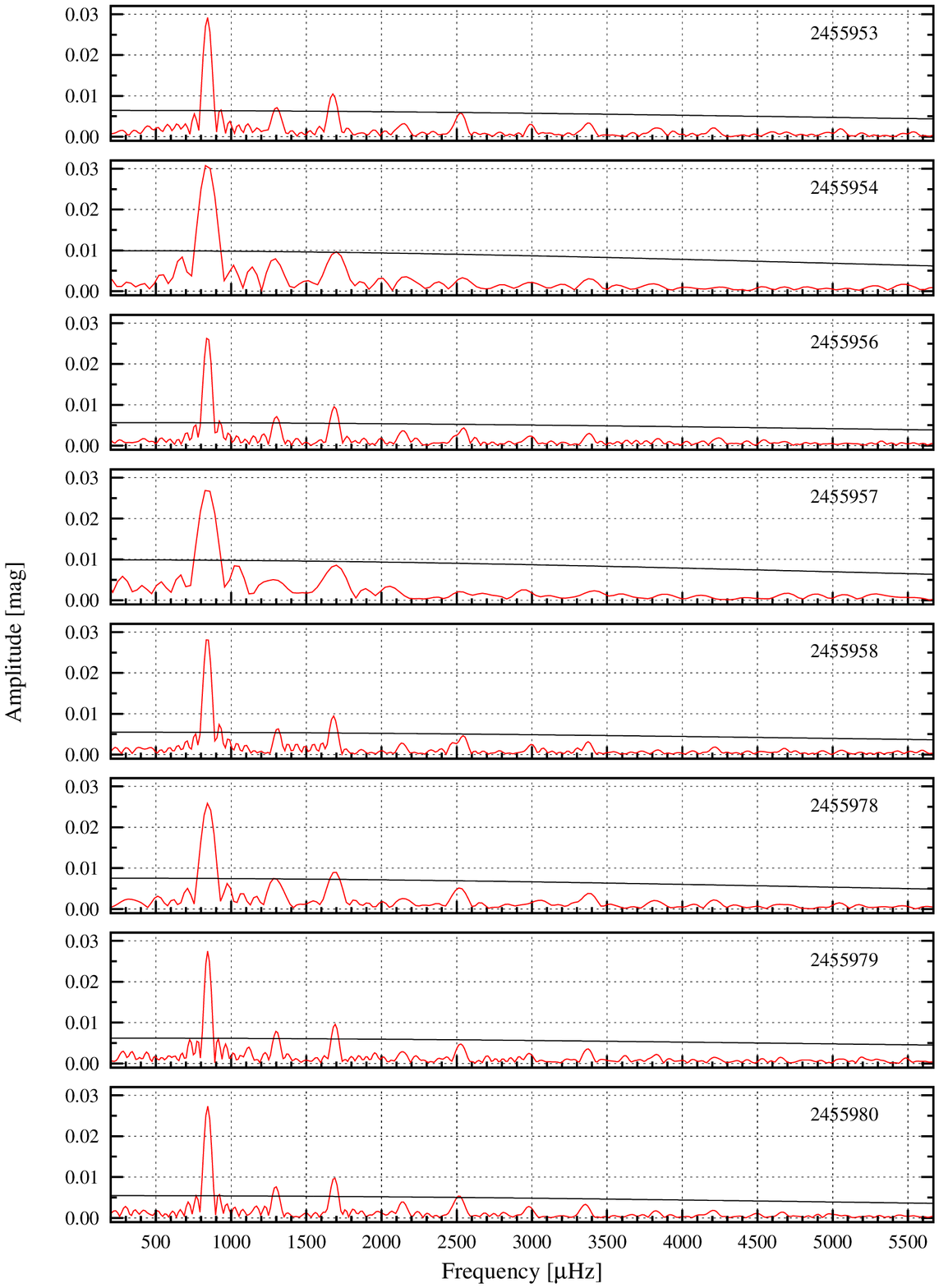}
\caption{Fourier transforms of the nightly observations of GD\,154 obtained at Konkoly Observatory. Black lines denote the 4$\langle {\rm A}\rangle$ significance levels.}{\label{app:gdsp}}
\end{figure*}

Figure~\ref{app:gdsp}: Fourier transforms of the nightly observations of GD\,154 obtained at Konkoly Observatory.

\section{}
%\label{app:wdft}

\begin{figure*}
\centering
\includegraphics[width=0.71\textwidth, angle=270]{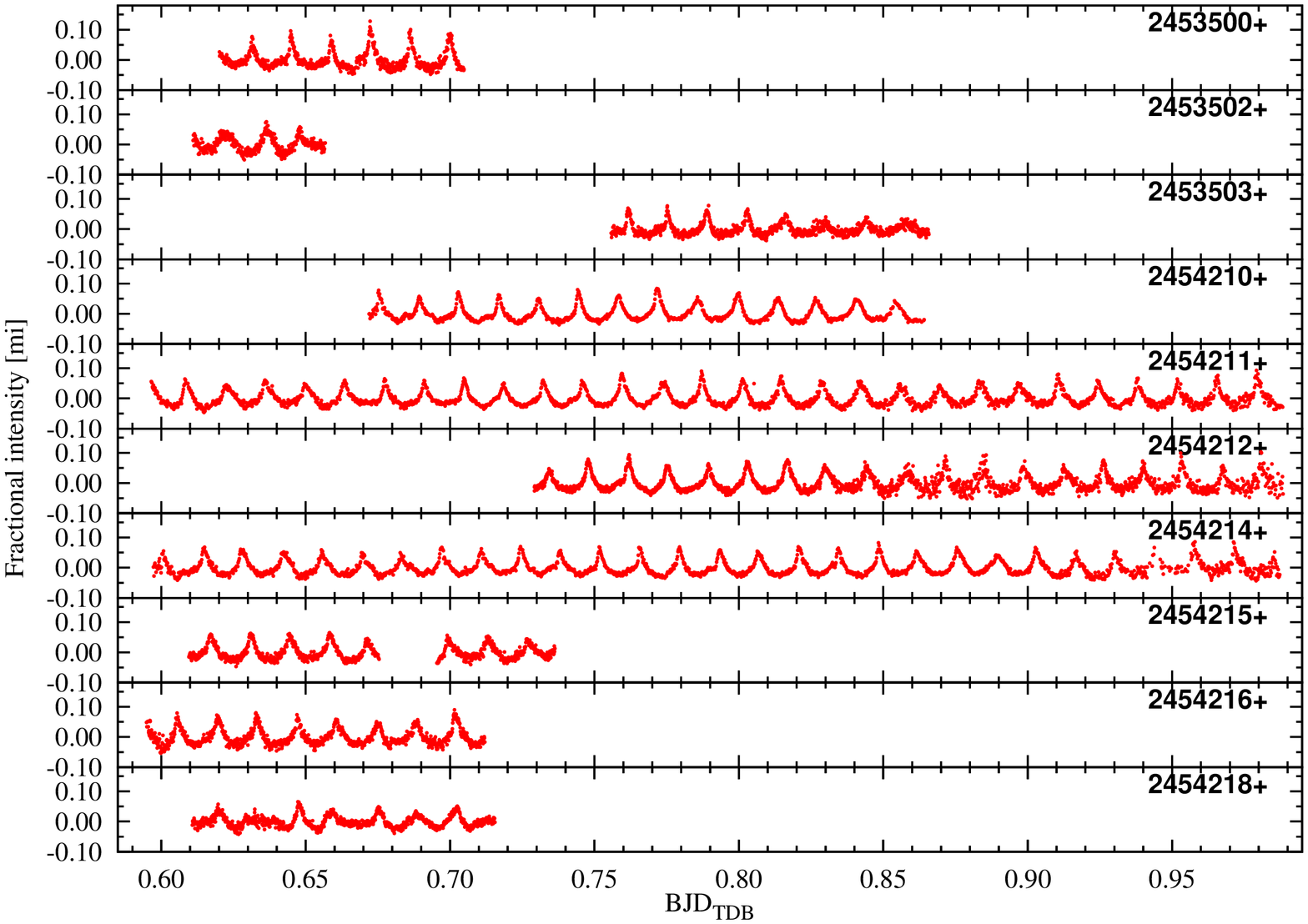}
\includegraphics[width=0.5\textwidth, angle=270]{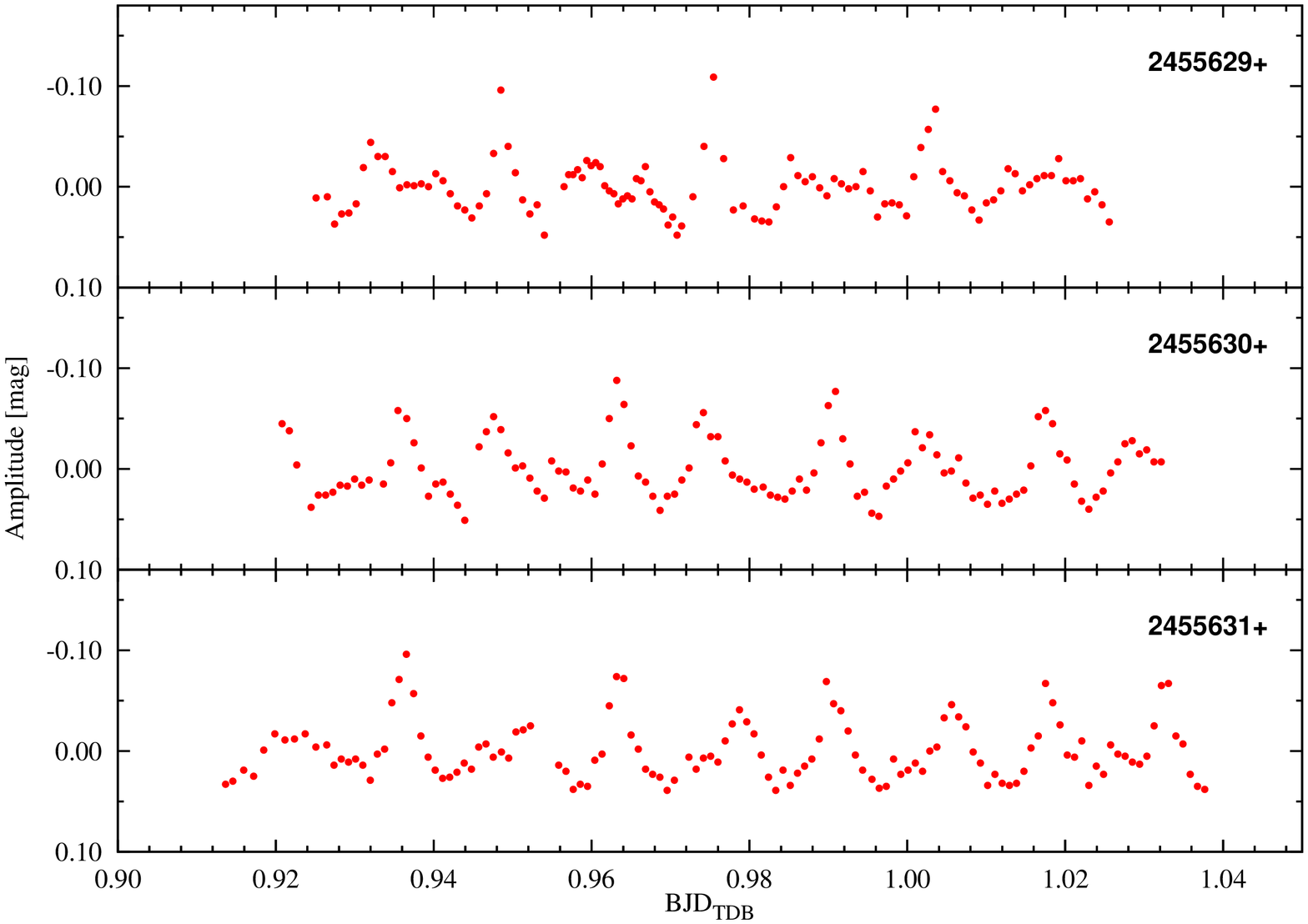}
\caption{Light curves of the observations of GD\,154 obtained at McDonald Observatory (top figure) and at Guillermo Haro Astrophysical Observatory (OAGH), Mexico (bottom figure), respectively.}{\label{app:gdlcusa}}
\end{figure*}

Figure~\ref{app:gdlcusa}: Light curves of the observations of GD\,154 obtained at McDonald Observatory (top figure) and at the National Astronomical Observatory, Mexico (bottom figure), respectively.

\section{}
%\label{app:wdft}

\begin{figure*}
\centering
\includegraphics[width=0.71\textwidth, angle=270]{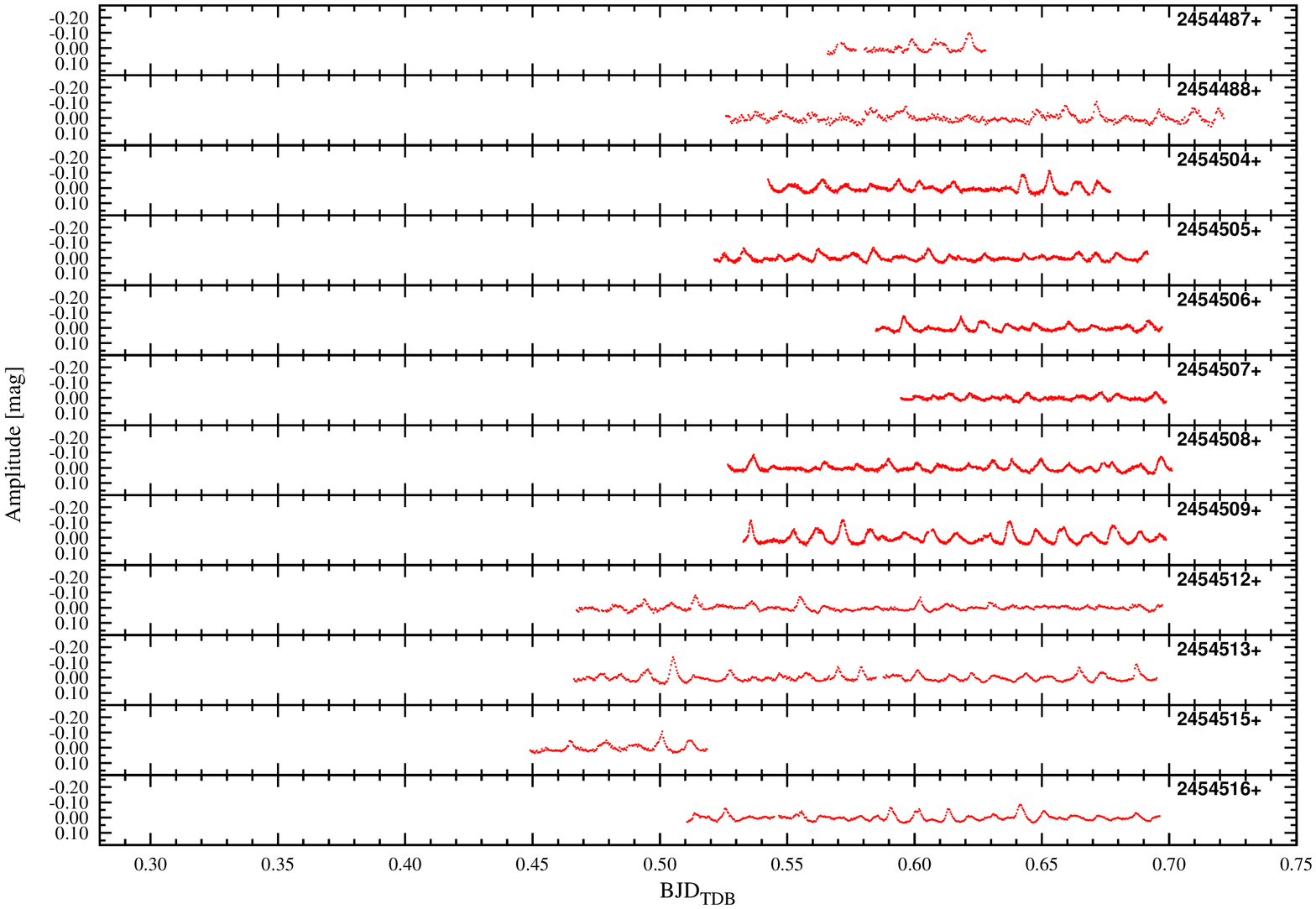}
\includegraphics[width=0.6\textwidth, angle=270]{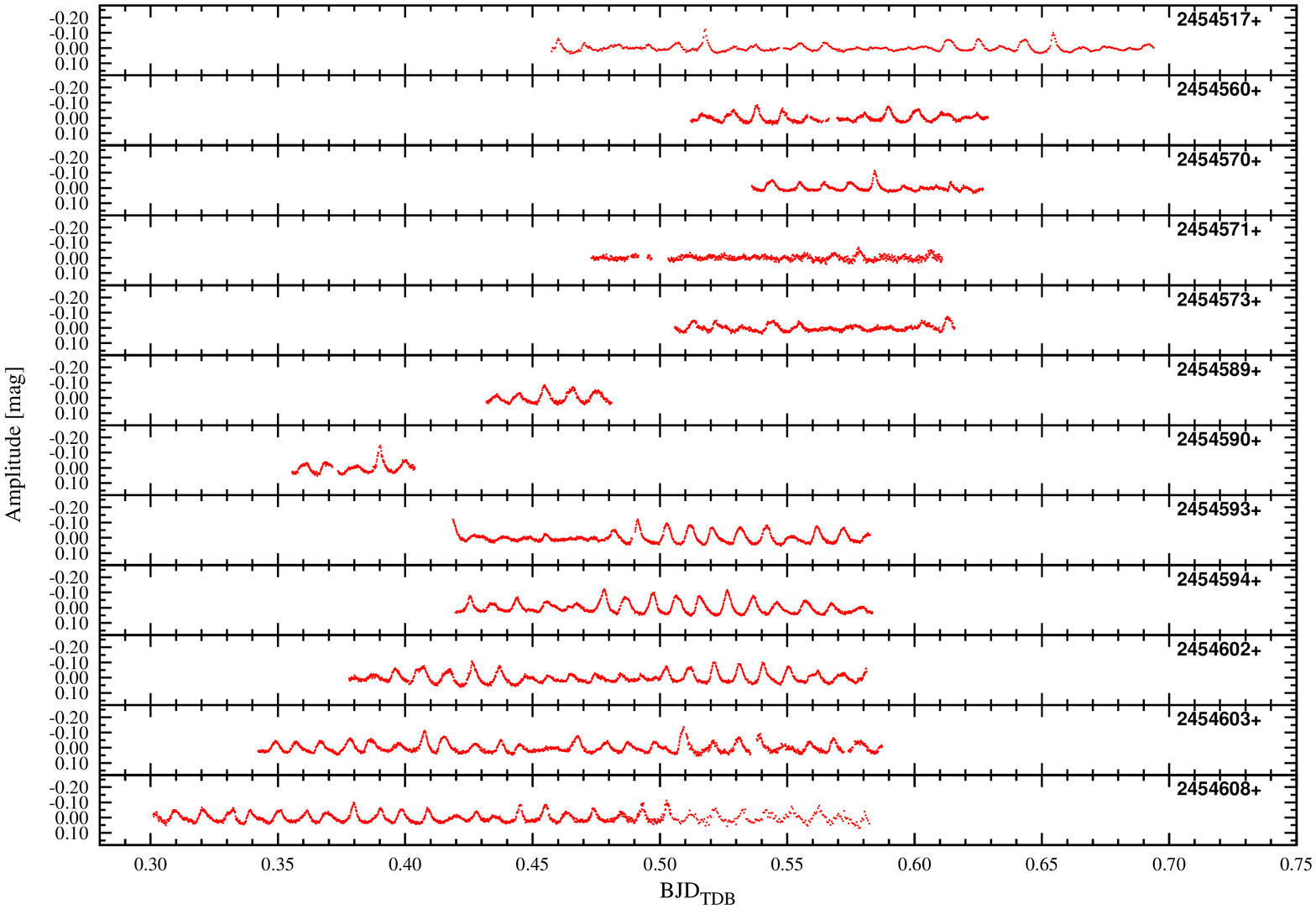}
\caption{Normalized differential light curves of the observations of R\,808.}{\label{app:rlc}}
\end{figure*}

Figure~\ref{app:rlc}: Normalized differential light curves of the observations of R\,808.

%%%%%%%%%%%%%%%%%%%%%%%%%%%%%%%%%%%%%%%%%%%%%%%%%%

% Don't change these lines
\bsp	% typesetting comment
\label{lastpage}
\end{document}